# A Novel Method for Scalable VLSI Implementation of Hyperbolic Tangent Function

Mahesh Chandra
NXP Semiconductors, India
mahesh.chandra_1@nxp.com

*Abstract*—Hyperbolic tangent and Sigmoid functions are used as non-linear activation units in the artificial and deep neural networks. Since, these networks are computationally expensive, customized accelerators are designed for achieving the required performance at lower cost and power. The activation function and MAC units are the key building blocks of these neural networks. A low complexity and accurate hardware implementation of the activation function is required to meet the performance and area targets of such neural network accelerators. Moreover, a scalable implementation is required as the recent studies show that the DNNs may use different precision in different layers. This paper presents a novel method based on trigonometric expansion properties of the hyperbolic function for hardware implementation which can be easily tuned for different accuracy and precision requirements.

## I. Introduction

Artificial neural networks (ANNs) have been used for modeling the complex non-linear relationships between the inputs and outputs in multiple applications. An ANN consists of a layered network of the artificial neurons which compute the weighted sum of multiple inputs and pass it through a non-linear activation function. State of the art deep neural networks (DNNs) have many such layers connected in feed forward fashion. Using these feed-forward DNNs, state of the art result has been achieved in various applications such as object detection and classification. However, there is another set of applications which requires the neural networks to model the history or sequence such as the natural language processing, classification of video sequences, and image captioning etc. Recurrent neural networks (RNNs) and long short-term memory (LSTM) have been used for such applications. These neural networks continue to use tanh activation function for its ability to handle vanishing gradients and ease of computing gradient.

Since, these algorithms require huge computing resources; there has been an effort to implement dedicated accelerators to speed up the execution. Activation function is one of the key building block required for the efficient hardware accelerator. Experimental study has shown that the accuracy of the activation function impacts the performance and the size of the neural networks. Hyperbolic tangent function, being a non-linear, function requires specific consideration for the accuracy and area trade-off. This paper presents a novel method for hardware implementation which can be easily tuned for different accuracy requirements.

## II. Literature Review

Tanh function, shown in figure 1, is a non-linear function defined as:

$$tanh(x) = \frac{e^x - e^{-x}}{e^x + e^{-x}} \qquad (1)$$

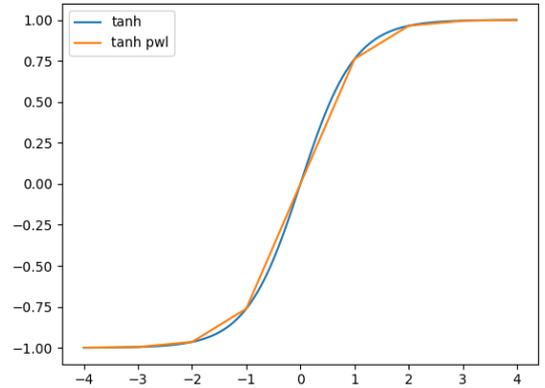

Fig. 1. tanh function and its piecewise linear approximation

Multiple implementations of hyperbolic tangent have been published in literature ranging from the simplest step and linear approximations to more complex interpolation schemes. This section reviews some of these methods for tanh implementation and discusses the motivation behind this paper.

The simplest implementation is to store the values of the function in a lookup table (LUT) and approximate the output with the lookup table value for the nearest input. Since, the function is non-uniform, it's challenging to balance the tradeoff between accuracy and area if the range is divided in equal sub-ranges. To address this issue, range addressable lookup table has been proposed by Leboeuf et al. [1]. The step size is varied depending on the variability of the function to reduce the size of LUT without impacting the accuracy. Another variation of this method is to use a two-step LUT. The first one with the coarse estimation and second one with the finer estimation. Namin et al. use this method but instead of an LUT, they use a combination of linear and saturation values for coarse approximation [2].

Zamanlooy et al. take advantage of the tanh function being an odd function and divide it in three ranges based on its basic properties; pass region, processing region and saturation region [3]. Then the hardware is optimized specific to the regions. In the pass region, the data is simply shifted and in the saturation region it is constant. In processing region, data is mapped from the input by simple bit-level mapping (i.e. the combinatorial logic).

The function value can be interpolated by piecewise linear (PWL) interpolation to reduce the error. The function value is stored in an LUT for known values and from these values, the function is interpolated for intermediate input values [4].

Adnan et al. have approximated the tanh function by Taylor series expansion [5]. The accuracy varies across the range of the input and the function is more accurately computed for smaller values of inputs. Moreover, if the number of terms in Taylor series are increased from three to four, improvement is just 2x where the error was large while it is 10x where the error was already small.

Abdelsalam et al. have used the DCT (discrete cosine transform) interpolation filter (DCTIF) for tanh approximation [6]. Like [3], they also divide the tanh function in three regions and use DCTIF for approximation in processing region. This method achieves higher accuracy than any of the published methods. However, it requires huge memory for storing the coefficients.

Rational interpolation methods have also been explored by the researchers. Z. Hajduk [7] has discussed the hardware implementation of tanh using Padé Approximant. Similalrly, Lambert's continuous fraction is also used for the rational function approximation of hyperbolic tangent. [8]

It is evident from this short list that various implementations have been published in the literature. Some of them are too complex and require huge resources and may be overkill for applications which work with fixed point data such as deep learning. Even though, rational approximations are computationally complex as they require a divider, they are worth exploring for proper comparison. Newton-Raphson method can be applied for the reciprocal computation to implement the divider [9]. Moreover, there are methods for fast oral calculation of various trigonometric and exponential function such as the one published by Ron Doerfler [10]. This method is quite interesting and can be applied to the hardware implementations. This paper explores a hardware implementation by adopting this method and making necessary changes to make it hardware friendly.

## III. METHOD OVERVIEW

This section discusses the method published by Ron Doerfler for oral approximation of Hyperbolic tangent function [10]. It basically consists of method of finding the hyperbolic tangent value for sum of two angles given the value of the function for two angles independently or if one of the angles is very small.

Hyperbolic tangent for the addition of two angles is given by:

$$\tanh(a+b) = \frac{\tanh a + \tanh b}{1 + \tanh a \times \tanh b} \quad (2)$$

Given tanh value at $a$, and very small $b$, it can be approximated as below:

$$\tanh b = b$$

$$\tanh(a+b) = \frac{\tanh a + b}{1 + b \times \tanh a}$$

$$\tanh(a+b) = (\tanh a + b) \times (1 - b \times \tanh a)$$

$$\tanh(a+b) = \tanh a + b \times (1 - \tanh^2 a) \quad (3)$$

This approximation works well for small '$b$'. For larger values of $b$, we can directly use equation (2). However, it requires operations which are costly for hardware or software implementation and difficult to parallelize. An alternative representation, that simplifies these operations, is presented in [10] and reproduced below.

Instead of working with tanh values, the author proposes to work with a transformed value called as velocity factor ($f$), and defined as:

$$f_a = \frac{1 + \tanh a}{1 - \tanh a} \quad (4)$$

To compute tanh from $f$, we can use following equation.

$$\tanh a = \frac{f_a - 1}{f_a + 1} \quad (5)$$

Given $f_a$ and $f_b$, $f_{a+b}$ can be computed as:

$$f_{a+b} = \frac{1 + \tanh(a+b)}{1 - \tanh(a+b)}$$

$$f_{a+b} = \frac{1 + \frac{\tanh a + \tanh b}{1 + \tanh a \times \tanh b}}{1 - \frac{\tanh a + \tanh b}{1 + \tanh a \times \tanh b}}$$

$$f_{a+b} = \frac{1 + \tanh a + \tanh b + \tanh a \times \tanh b}{1 - \tanh a - \tanh b + \tanh a \times \tanh b}$$

$$f_{a+b} = \frac{(1 + \tanh a) \times (1 + \tanh b)}{(1 - \tanh a) \times (1 - \tanh b)}$$

$$f_{a+b} = f_a \times f_b \quad (6)$$

Given this velocity factor for sum of angles, hyperbolic function can be computed back using (5).

## IV. HARDWARE IMPLEMENTATION

Since, tanh is an odd function, the main algorithm can be implemented for positive values only. Main steps as shown in figure 2, are sign detection, tanh value computation and sign conversion. Computing tanh only for positive values simplifies the hardware implementation; hence all the analysis in subsequent sections considers the tanh computation for positive input values only.

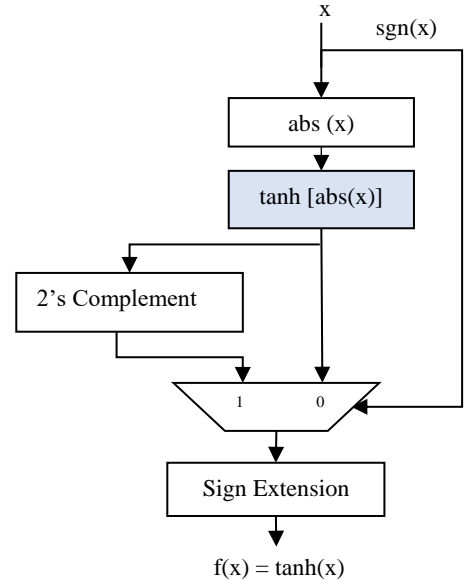

Fig. 2 . Data Flow diagram for tanh function implementation

It has been shown that the inference using DNNs is less sensitive to the quantization of the data (i.e. the precision); so, the data is assumed to be organized as 16-bit or 8-bit fixed point signed input to the tanh for the discussion in this paper.

For 16-bit fixed point input data, we can consider 13-bit or 12-bit precision for fractional part. The range of the input data in this case will be either (-4,4) or (-8,8) respectively. However, for practical purposes, we can constrain the domain to $\tanh^{-1}[\pm(1-2^{-b})]$ where $b$ is the number bits used to represent the fractional part of tanh at the output. For 8, 12 and 16-bit signed fixed-point representation with fractional only, the corresponding domain is ±2.77 (±2.42), ±4.16 (±3.82) and ±5.55 (±5.20) respectively. Beyond this domain, the errors for tanh is smaller than that can be represented by the least significant bit and can be ignored. For the following discussion, the maximum error is restricted to the lsb.

*A. Hardware Implementation of Published Method*

The manipulations described in (3)-(6) make the hardware implementation less complex compared to direct computation using (2). For example, if there is an n-bit integer N represented as $b_{n-1}b_{n-2}..b_2b_1b_0$ in binary, then velocity factor $f_N$ can be computed as:

$$f_N = \prod_{k=0}^{n-1} f_{2^{b_k \times k}} \quad (7)$$

Once $f_N$ is known, tanh can be computed using (5).

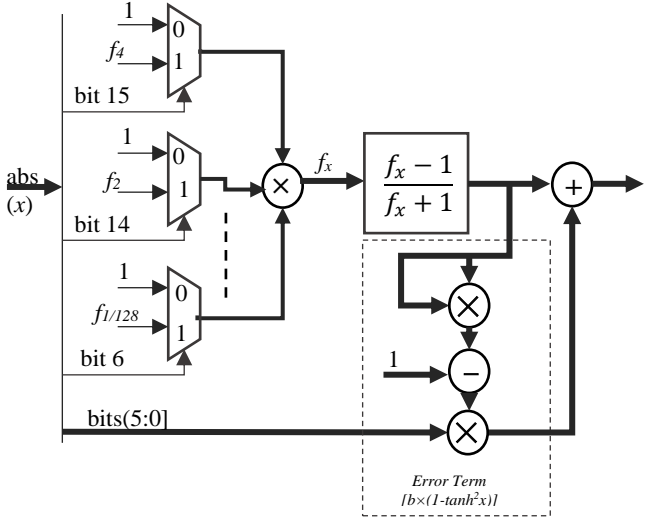

Fig. 3. High level Block diagram for tanh commputation using published method for input represneted in s3.12 format

For hardware implementation, shown in figure 3, velocity factors can be stored in registers for the numbers which are the power of two and more than a threshold (e.g. $2^{-7}$). Equations (7) and (5) are used to compute the tanh value for the sum of stored numbers. For the addition smaller than threshold, equation (3) can be used for compensating the error. The value of this threshold affects the accuracy of the approximation and the number of registers required for storing velocity factors. The simplest implementation using this method for s3.12 input requires 10 registers storing the tanh velocity factor ($f_a$) value for $2^k$ (-7 ≤ k ≤ 2), and 9 multipliers (one for each bit).

The implementation requires a division operation which can be implemented by multiplying numerator with the reciprocal of denominator. Newton Raphson method can be used along with some data manipulation techniques for computing reciprocal [11]. The data manipulation is required to bring the denominator in the range of (0.5,1) required by the method. Newton Raphson method iteratively refines the initial guess $x_0$ for reciprocal of a number '$b$' using equation (8). The high-level block diagram for Newton Raphson method for computing reciprocal is shown in figure 4.

$$x_{i+1} = x_i \times (2 - b \times x_i) \quad (8)$$

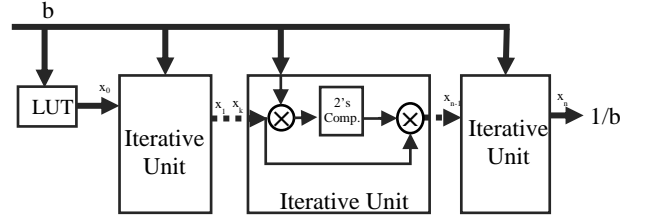

Fig. 4. High level block diagram of iterative Newton Raphson method for reciprocal computation

*B. Architectural Improvements for Scalable Hardware Implementation*

*1) Removing the last stage multiplier*

The equation (3) used in original approximation and shown in dotted rectangular block in figure 3, has two main problems; one, it introduces error and two, it requires two multipliers in the last stage. We can get rid of this and instead use (7) for computing velocity factor for all bit positions instead of only some MSBs. It makes the tanh computation highly accurate and the error is introduced only because of the precision of the numbers and arithmetic. So, for our implementation trials we can make this modification i.e. store the velocity factor for all bit positions and no approximation. This also requires additional multipliers for computing (7); however, this can be optimized by hardware manipulations and is discussed later.

*2) Velocity Factor Range and Precision*

For 16-bit fixed point data in s3.12 format, the velocity factor range is [1.0004884,54.59815]. It requires at least 6.11 bits to represent it. This dynamic range is a function of the input precision and range and hence makes the scalability a bit difficult to handle. It would be nice to have the fractional range that can be represented as 0.N and select N based on the input and output precision. Fortunately, such a representation is possible by reworking the algebraic manipulations.

Instead of storing tanh values in the LUTs, redefine the velocity factor $f$ as:

$$f_a = \frac{1-\tanh a}{1+\tanh a} \quad (9)$$

To compute tanh from $f$, we can use following equation.

$$\tanh a = \frac{1-f_a}{1+f_a} \quad (10)$$

There is no change for computing $f_{a+b}$; and given $f_a$ and $f_b$, (6) can be used for this purpose. Using this method, the $f_a$ is always in the range of (0,1) and makes the implementation more friendly to the scaling.

This has another advantage in the implementation of the division logic using the Newton Raphson method. The Newton Raphson method works well if the operand is in the range of (0.5,1). Since,

$$f_a \in (0,1) \Rightarrow (1+f_a) \in (1,2) \Rightarrow \frac{(1+f_a)}{2} \in (0.5,1) \quad (11)$$

So, a single right shift brings the denominator in the required range.

### 3) Reducing Number of Multipliers in Computing Velocity Factor

We can also reduce the number of multipliers by using LUTs instead of registers for storing velocity factors. Instead of storing values of velocity factors for a single place value, we can store the velocity factor corresponding to combination of them. For example, we can combine two bits and store the values as given by table I. This reduces number of multipliers at the cost of LUT which store constant values and can be optimized. If we store LUT entries for four bits together, number of multipliers reduced to 3 for s3.12 representation at the cost of 4 LUTs each storing 16 velocity factor values.

TABLE I. MULTI-BIT LOOKUP FOR VELOCITY FACTORS

| Bits | Value |
|---|---|
| 00 | 1.0 |
| 01 | Velocity factor corresponding to lsb |
| 10 | Velocity factor corresponding to msb |
| 11 | Multiplication of velocity factors corresponding to lsb and msb |

Using LUTs in this way adds new challenges. Since, velocity factor values are fractions, their multiplication results in even smaller numbers and requires higher number of bits to represent them and preserve the precision. If we simply combine them in increasing order e.g. LUT0 storing the velocity factors for $2^{-12}$, $2^{-11}$, $2^{-10}$, $2^{-9}$; then, the problem is accentuated. Instead, we can combine the bit positions differently depending on magnitude to reduce the impact; for example, LUT0 can store the velocity factors for $2^{-12}$, $2^{-5}$, $2^{-4}$, $2^{2}$. What this means for hardware is that instead of using bits $\{x_3, x_2, x_1, x_0\}$ as address for LUT0, we chose $\{x_{15}, x_8, x_7, x_0\}$ bits as the address for LUT0. With this scheme of addressing, 18-bit precision is enough for the 1-bit error on the output. Note that the bit shuffling doesn't add any hardware cost.

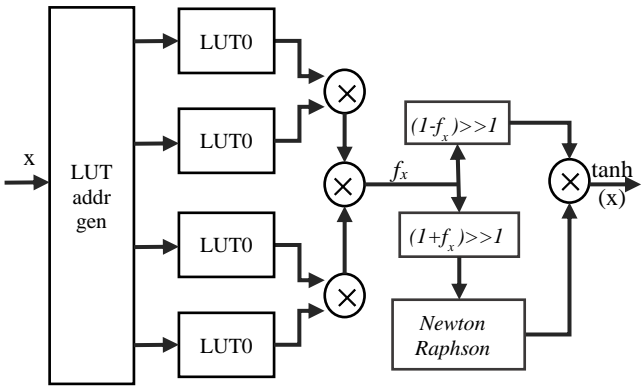

Fig. 5. High level Block diagram of optimized HW for tanh computation

### 4) Adder, Subtractor and Divider

The last stage of the tanh computation requires a subtractor $(1-f_x)$, an adder $(1+f_x)$ and a divider $((1-f_x)/(1+f_x))$. The divider can be implemented using three-stage Newton Raphson method and a multiplier. Since, $f_x$ is in the range (0,1), adding one is simply bit concatenation (i.e. suffix) for hardware and no real adder is required. The subtractor is a 2's complement logic and can be approximated by 1's complement without introducing much error as discussed later.

### 5) Putting Together

These optimizations result in significantly simpler hardware architecture as shown in fig. 5.

## V. IMPLEMENTATION RESULTS

The method discussed above is highly accurate method and the error is introduced due to precision and few approximations such as Newton Raphson for reciprocal computation and 1's complement for subtraction instead of 2's complement. The Table II summarizes the error introduced due to these approximations for s3.12 input and s.15 output. The precision of LUTs and multipliers is kept at 18 and 16 bits respectively. It's evident from the table that Newton Raphson method with 1's complement subtraction gives as good accuracy as real divider. Similarly, using 1's complement for $(1-f_x)$ computation drops the accuracy marginally to $5.87 \times 10^{-5}$ from $4.32 \times 10^{-5}$.

TABLE II. ERROR ANALYSIS FOR ARITHMETIC APPROXIMATIONS

| No. of Newton Raphson Iteration Stages | Subtractor | Max Error |
|---|---|---|
| 0 (Floating Point Divider followed by fixed point conversion for reference) | - | $4.44 \times 10^{-5}$ |
| 2 | 1's | $2.77 \times 10^{-4}$ |
| 2 | 2's | $2.56 \times 10^{-4}$ |
| 3 | 1's | $4.32 \times 10^{-5}$ |
| 3 | 2's | $4.44 \times 10^{-5}$ |

A reusable RTL code was written using Verilog HDL and synthesized for PPA (power, performance and area) trade-off analysis. The precision of input and output can be controlled by parameters selecting the bit width of LUTs and multipliers etc. The PPA trade-offs require multiple pipelined designs. The performance for different pipeline stages and 16- and 8-bit input are given in table III and IV.

TABLE III. SUMMARY OF RESULTS FOR DIFFERENT FLAVOURS OF TANH IMPLEMENTATION FOR S3.12 INPUT AND S.15 OUTPUT

| Cells | Latency (Clocks) | Area (um²) | Leakage Power (uW) | Max Frequency (MHz) | Logic Levels |
|---|---|---|---|---|---|
| SVT | 1 | 3748.28 | 4.20 | 188 | 135 |
| LVT | 1 | 2600.34 | 119.33 | 302 | 111 |
| SVT | 2 | 3400.43 | 3.53 | 258 | 95 |
| LVT | 2 | 3367.16 | 180.67 | 511 | 86 |
| SVT | 7 | 3688.98 | 3.92 | 1176 | 25 |
| LVT | 7 | 3147.68 | 146.67 | 2134 | 17 |

TABLE IV. SUMMARY OF RESULTS FOR DIFFERENT FLAVOURS OF TANH IMPLEMENTATION FOR S3.5 INPUT AND S.7 OUTPUT

| Cells | Latency (Clocks) | Area (um²) | Leakage Power (uW) | Max Frequency (MHz) | Logic Levels |
|---|---|---|---|---|---|
| SVT | 1 | 764.37 | 0.81 | 254 | 97 |
| LVT | 1 | 568.99 | 24.19 | 303 | 95 |
| SVT | 2 | 885.29 | 0.99 | 364 | 74 |
| LVT | 2 | 877.82 | 51.67 | 715 | 70 |
| SVT | 7 | 995.60 | 1.08 | 1532 | 14 |
| LVT | 7 | 934.82 | 49.04 | 2985 | 13 |

As discussed earlier, PWL and Taylor series expansion are quite popular for non-linear function implementation. However, they suffer from lack of scalability as the LUT size or number of terms must change as the accuracy requirement changes. DCT interpolation technique offers high accuracy but it requires huge memory for storing coefficients [6]. Other high accuracy implementations, such as using Padé approximants and CORDIC have higher latencies [7].

## VI. Conclusion

This paper presents the hardware implementation of a highly accurate method for tanh computation. Though, the method itself is error free as against the series approximation or the PWL; the limited precision and reciprocal approximation introduce some error. The implementation presented here offers a highly accurate and scalable circuit for tanh computation.